\documentclass[twocolumn,showpacs,superscriptaddress]{revtex4b4}

\usepackage{graphicx}
\newcommand{\nft}{$^{15}{\rm ND}_3$\ }
\newcommand{\gen}{$G_E^n\ $}

\begin{document} \title{A Measurement of the Electric Form Factor of the Neutron
\\through $\vec{d}(\vec{e},e'n)p$ at $Q^2 = 0.5 $ (GeV/c)$^2$}

\date{March 26, 2001}

\author{H.~Zhu}\affiliation{Institute of Nuclear and Particle Physics, University of Virginia, Charlottesville, VA 22903}
\author{A.~Ahmidouch}\affiliation{North Carolina A\&T State University, Greensboro, NC 27411}
\affiliation{Thomas Jefferson National Accelerator Facility, Newport News, VA 23606}
\author{H.~Anklin}\affiliation{Florida International University, Miami, FL 33199}\affiliation{Thomas Jefferson National Accelerator Facility, Newport News, VA 23606}
\author{H. Arenh\"{o}vel}\affiliation{Institut f\"{u}r Kernphysik, Johannes
Gutenberg-Universit\"{a}t, D-55099 Mainz, Germany}
\author{C.~Armstrong} \affiliation{Thomas Jefferson National Accelerator Facility, Newport News, VA 23606}
\author{C.~Bernet}\affiliation{Universit\"{a}t Basel, CH-4056  Basel, Switzerland}
\author{W.~Boeglin}\affiliation{Florida International University, Miami, FL 33199}\affiliation{Thomas Jefferson National Accelerator Facility, Newport News, VA 23606}
\author{H.~Breuer}\affiliation{University of Maryland, College Park, MD 20742}
\author{P.~Brindza}\affiliation{Thomas Jefferson National Accelerator Facility, Newport News, VA 23606}
\author{D.~Brown}\affiliation{University of Maryland, College Park, MD 20742}
\author{S.~B\"{u}ltmann} \affiliation{Institute of Nuclear and Particle Physics, University of Virginia, Charlottesville, VA 22903}
\author{R.~Carlini}\affiliation{Thomas Jefferson National Accelerator Facility, Newport News, VA 23606}
\author{N.~Chant}\affiliation{University of Maryland, College Park, MD 20742}
\author{A.~Cowley}\altaffiliation{Permanent Address: Department of Physics,
 University of Stellenbosch, Private Bag X1, Matieland, 7602, South Africa}
\affiliation{University of Maryland, College Park, MD 20742}\affiliation{Thomas Jefferson National Accelerator Facility, Newport News, VA 23606}
\author{D.~Crabb}\affiliation{Institute of Nuclear and Particle Physics, University of Virginia, Charlottesville, VA 22903}
\author{S.~Danagoulian}\affiliation{North Carolina A\&T State University, Greensboro, NC 27411}\affiliation{Thomas Jefferson National Accelerator Facility, Newport News, VA 23606}
\author{D.B.~Day}\affiliation{Institute of Nuclear and Particle Physics, University of Virginia, Charlottesville, VA 22903}
\author{T.~Eden}\affiliation{Norfolk State University, Norfolk, VA 23504}
\author{R.~Ent}\affiliation{Thomas Jefferson National Accelerator Facility, Newport News, VA 23606}
\author{Y.~Farah}\affiliation{North Carolina A\&T State University, Greensboro, NC 27411}
\author{R.~Fatemi}\affiliation{Institute of Nuclear and Particle Physics, University of Virginia, Charlottesville, VA 22903}
\author{K.~Garrow}\affiliation{Thomas Jefferson National Accelerator Facility, Newport News, VA 23606}
\author{C.~Harris}\affiliation{Institute of Nuclear and Particle Physics, University of Virginia, Charlottesville, VA 22903}
\author{M.~Hauger}\affiliation{Universit\"{a}t Basel, CH-4056  Basel, Switzerland}
\author{A.~Honegger}\affiliation{Universit\"{a}t Basel, CH-4056  Basel, Switzerland}
\author{J.~Jourdan}\affiliation{Universit\"{a}t Basel, CH-4056  Basel, Switzerland}
\author{M.~Kaufmann}\affiliation{Universit\"{a}t Basel, CH-4056  Basel, Switzerland}
\author{M.~Khandaker}\affiliation{Norfolk State University, Norfolk, VA 23504}
\author{G.~Kubon}\affiliation{Universit\"{a}t Basel, CH-4056  Basel, Switzerland}
\author{J.~Lichtenstadt}\affiliation{School of Physics and Astronomy, Tel Aviv University, Tel Aviv, 69978 Israel}
\author{R.~Lindgren}\affiliation{Institute of Nuclear and Particle Physics, University of Virginia, Charlottesville, VA 22903}
\author{R.~Lourie}\affiliation{State University of New York, Stonybrook, NY 11794}
\author{A.~Lung}\affiliation{Thomas Jefferson National Accelerator Facility, Newport News, VA 23606}
\author{D.~Mack}\affiliation{Thomas Jefferson National Accelerator Facility, Newport News, VA 23606}
\author{S.~Malik}\affiliation{School of Physics and Astronomy, Tel Aviv University, Tel Aviv, 69978 Israel}
\author{P.~Markowitz}\affiliation{Florida International University, Miami, FL 33199} \affiliation{Thomas Jefferson National Accelerator Facility, Newport News, VA 23606}
\author{K.~McFarlane}\affiliation{Norfolk State University, Norfolk, VA 23504}
\author{P.~McKee}\affiliation{Institute of Nuclear and Particle Physics, University of Virginia, Charlottesville, VA 22903}
\author{D.~McNulty}\affiliation{Institute of Nuclear and Particle Physics, University of Virginia, Charlottesville, VA 22903}
\author{G.~Milanovich}\affiliation{Institute of Nuclear and Particle Physics, University of Virginia, Charlottesville, VA 22903}
\author{J.~Mitchell}\affiliation{Thomas Jefferson National Accelerator Facility, Newport News, VA 23606}
\author{H.~Mkrtchyan}\affiliation{Yerevan Physics Institute, Yerevan, Armenia}
\author{M.~M\"{u}hlbauer}\affiliation{Universit\"{a}t Basel, CH-4056  Basel, Switzerland}
\author{T.~Petitjean}\affiliation{Universit\"{a}t Basel, CH-4056  Basel, Switzerland}
\author{Y.~Prok}\affiliation{Institute of Nuclear and Particle Physics, University of Virginia, Charlottesville, VA 22903}
\author{D.~Rohe}\affiliation{Universit\"{a}t Basel, CH-4056  Basel, Switzerland}
\author{E.~Rollinde}\affiliation{University of Maryland, College Park, MD 20742}
\author{O.A.~Rondon}\affiliation{Institute of Nuclear and Particle Physics, University of Virginia, Charlottesville, VA 22903}
\author{P.~Roos}\affiliation{University of Maryland, College Park, MD 20742}
\author{R.~Sawafta}\affiliation{North Carolina A\&T State University, Greensboro, NC 27411}
\author{I.~Sick}\affiliation{Universit\"{a}t Basel, CH-4056  Basel, Switzerland}
\author{C.~Smith}\affiliation{Institute of Nuclear and Particle Physics, University of Virginia, Charlottesville, VA 22903}
\author{T.~Southern}\affiliation{North Carolina A\&T State University, Greensboro, NC 27411}
\author{M.~Steinacher}\affiliation{Universit\"{a}t Basel, CH-4056  Basel, Switzerland}
\author{S.~Stepanyan}\affiliation{Yerevan Physics Institute, Yerevan, Armenia}
\author{V.~Tadevosyan}\affiliation{Yerevan Physics Institute, Yerevan, Armenia}
\author{R.~Tieulent}\affiliation{University of Maryland, College Park, MD 20742}
\author{A.~Tobias}\affiliation{Institute of Nuclear and Particle Physics, University of Virginia, Charlottesville, VA 22903}
\author{W.~Vulcan}\affiliation{Thomas Jefferson National Accelerator Facility, Newport News, VA 23606}
\author{G.~Warren}\affiliation{Universit\"{a}t Basel, CH-4056  Basel, Switzerland}
\author{H.~W\"{o}hrle}\affiliation{Universit\"{a}t Basel, CH-4056 Basel, Switzerland}
\author{S.~Wood}\affiliation{Thomas Jefferson National Accelerator Facility, Newport News, VA 23606}
\author{C.~Yan}\affiliation{Thomas Jefferson National Accelerator Facility, Newport News, VA 23606}
\author{M.~Zeier}\affiliation{Institute of Nuclear and Particle Physics, University of Virginia, Charlottesville, VA 22903}
\author{J.~Zhao}\affiliation{Universit\"{a}t Basel, CH-4056  Basel, Switzerland}
\author{B.~Zihlmann}\affiliation{Institute of Nuclear and Particle Physics, University of Virginia, Charlottesville, VA 22903}¬
\begin{abstract} We report the first measurement of the neutron electric form factor \gen
 via $\vec{d}(\vec{e},e'n)p$ using a solid polarized target. \gen was
determined from the beam--target asymmetry in the scattering of longitudinally 
polarized electrons from polarized deuterated ammonia (\nft). 
The measurement  was performed in Hall C at Thomas Jefferson National Accelerator
Facility (TJNAF) in  
quasi free kinematics with the target polarization perpendicular to the momentum 
transfer.
The electrons were detected in  
a magnetic spectrometer in coincidence with neutrons in a large solid angle 
segmented  detector.
We find $G_E^n =  0.04632\pm0.00616 ({\rm stat.}) \pm0.00341 ({\rm syst.})$ at $Q^2 = 0.495$ (GeV/c)$^2$. 
\end{abstract}
\maketitle
Precise data on the neutron (and proton) form factors
are important for understanding the non--perturbative mechanism responsible
for confinement and are necessary in the interpretation of the electromagnetic
properties of nuclei. The magnetic form factor of the
neutron \cite{An94} has recently been measured with
high precision. The neutron charge form factor \gen, in contrast, is only now yielding to intense
efforts focused on its determination.

The major difficulty faced in a measurement of the neutron form  factors is  the
lack of a free neutron target. The determination of \gen is further impeded by
its small  size.  
Advances in  polarized electron sources, CW accelerators, polarimeters
and polarized targets, now allow $G_E^n$ to be extracted from  experiments which
 exploit spin degrees of freedom. In particular the interference of the
magnetic and electric
scattering amplitudes is  responsible for an
asymmetry that
can be measured in both polarized electron/polarized target  experiments ($\vec{d}(\vec{e},e'n)p$  \cite{Pa99}
and $\vec{^3{\rm He}}(\vec{e},e'n)p$  \cite{Ro99,Be99,Me94}) and in polarized electron recoil
polarization measurements ($d(\vec{e},e'\vec{n})p$  \cite{Ed94,He99,Os99}).

For a vector polarized target of free neutrons, with
the polarization $P_n$, in the scattering plane and perpendicular
to the momentum transfer $\vec{q}$, \gen 
is related to the helicity 
asymmetry $A_{en}^V$  \cite{Do89} by

\begin{equation}A^V_{en} =
\frac{-2\sqrt{\tau(\tau+1)}
\tan(\theta_e/2)G_E^nG_M^n}{(G_E^n)^2 +
\tau[1+2(1+\tau)\tan^2(\theta_e/2)](G_M^n)^2 }\end{equation} where $Q^2$ is the
 four momentum transfer,  $\tau =
Q^2/4M_n^2$, and $\theta_e$ is the electron scattering angle.
$A^V_{en}$ is related to the experimental counts asymmetry $\epsilon = (L-R)/(L+R)$,
where $L$ and $R$ are charge normalized counts for opposite beam
polarizations $P_e$, by $A^V_{en} = \epsilon/(P_{e} P_{n} f)$,
 where $f$ is the dilution factor due to scattering from materials other than polarized neutrons.

In practice one measures the helicity asymmetry  from polarized
deuterons in quasi elastic kinematics.
 The counts  asymmetry  for polarized electron polarized deuteron scattering
 can be written  \cite{Ar88,Ar92} as:
 \begin{equation}
	\epsilon = f \frac{P_e A_e + P_e P_1^d A^V_{ed} + P_e
 P_2^d A^T_{ed}}{1 + P_1^d A^V_d + P_2^d A^T_d} 
\label{eq:sigma}\end{equation} 
where $P_{1(2)}^d$ is the target
vector (tensor) polarization, 
and $A_e,A_d^V, A_d^T, A_{ed}^V$, and $A_{ed}^T$ are the
electron beam induced asymmetry, the vector and tensor deuteron target
asymmetries, and the  deuteron vector and tensor beam--target asymmetries,
respectively. For experiments with the target polarization in the
scattering plane which sample the neutron Fermi cone in an
azimuthally symmetric way this reduces to
\begin{equation}
	\epsilon = f \frac{P_e P_1^d A^V_{ed}}{1 + P_2^d A^T_d}.
\end{equation}
Appropriate averaging of data where the the sign of $P_1^d$ is
reversed reduces the sensitivity to $A_e$ and $A_{ed}^T$ if
the setup does not have perfect azimuthal symmetry. 
For most practical targets $P_2^d$ is small ($~3 \% $) and the
second term in the denominator may be neglected. Realistic
calculations indicate that $A^V_{ed}$ has a linear sensitivity to
the magnitude of \gen for $d(e,e'n)$ at low recoil momentum \cite{Ar88,Ar92}.

We present in this letter a measurement at $Q^2 = 0.495$ (GeV/c)$^2$ carried out at TJNAF, where
the accelerator provides CW longitudinally polarized electrons. 
This experiment took place in Hall C with an incident electron
energy of 2.725 GeV and a beam current of $\approx$ 100~nA incident on a
dynamically polarized solid deuterated ammonia target. The scattered electrons were detected in
coincidence with the knockout neutrons.

The polarized electron beam was produced by photoemission from a strained-layer
semiconductor cathode illuminated by circularly polarized laser light 
 \cite{Si97} at the accelerator injector. 
The helicity of the beam was changed in a pair--wise pseudo-random sequence once per second
to minimize sensitivity to instrumental drifts. The longitudinal polarization of the
electrons was measured at regular intervals during the experiment 
 with a M\o ller polarimeter \cite{Hau00} just upstream
of the target. The average beam polarization for the
data taking was   $P_{e}=0.776 \pm 0.002$ (stat.).

To prevent localized heating of the target material and 
to insure uniform irradiation, 
the beam was rastered over the face of the target 
  such that the full face of the target was illuminated 
 during each helicity state.
The beam position was recorded by a secondary emission monitor
 \cite{steinacher00} consisting of thin stainless steel strips
in both the horizontal and vertical directions. It gave the
transverse position of the interaction point of the electron
and provided a calibration of the raster magnet
system so that the beam position at the target could be used in the reconstruction.

The polarized target \cite{DCrDBD} included a permeable target cell 
filled with granules of \nft submerged in
liquid He  maintained at 1K by a high power He
evaporation refrigerator. 
A  5~T magnetic field was provided by a
superconducting coil  arranged as a Helmholtz pair. 
The magnetic field was in the
horizontal plane  perpendicular to  $<\vec{q}>$  at 151.6$^\circ$ with respect to  the
beam direction.  The field orientation
was measured in situ with a Hall probe  to $\pm 0.1^\circ$.  A three magnet
chicane compensated for the effects of the target magnetic field on the incident
electrons. The field effects on the scattered electrons tilted
the scattering plane by 4$^\circ$ with respect to the horizontal plane. The
\nft had been previously irradiated
at either Stanford or at the TJNAF Free Electron Laser
with low energy electrons ($\simeq 35$ MeV) to
introduce a dilute collection of paramagnetic centers. 
The material was
polarized by the dynamic nuclear polarization method. 
The polarization was measured continuously via NMR using a series LCR circuit
and Q meter detector  \cite{Court93}. 
The average deuteron polarization throughout the experiment was $P_1^d=0.21 \pm 0.01$.

The High Momentum Spectrometer (HMS) in its standard configuration was set
at 15.7$^\circ$  to
 detect the scattered
electrons.
Modifications were made to the standard
reconstruction algorithm of the HMS to account for the target field and the
beam raster offset.

Knockout nucleons ($p$ and $n$) were detected in an array
of plastic scintillators. It consisted  of 2 planes of thin (0.6~cm)  veto
paddles and 5 planes of 10~cm thick bars. 
The scintillator bars (160 cm long in the horizontal direction)
had a phototube at each end to allow good position and timing resolution. The detector
was positioned  at 61.6$^\circ$ (along the direction of $<\vec{q}>$),   providing a
solid angle of $\approx  160$\  msr and was enclosed in a
thick concrete walled hut open 
towards the target. The front shielding consisted of  16.7  mm  of
lead and 25 mm of CH$_2$ sheets.
The time resolution was  450 ps ($\sigma$) as determined from the time of flight peak of photons (from
$\pi^0$ decay) in the meantime spectrum.
With the detector
positioned 4.0~m from the target for  nucleons with kinetic energy of 267 MeV
it provided an energy resolution of 16.5 MeV.
The neutron energy combined with the
scattered electron energy allowed us to
eliminate events associated with pion production.
The neutron vertical position  was determined by the
segmentation of the detector (10 cm) while the horizontal position 
was determined from the time difference of the
phototubes on the first bar hit along the $n$ track. The
measured horizontal resolution was $\approx$ 5 cm.

 The electron-nucleon trigger was formed by a
coincidence between the HMS electron and a hit in any one of the 5
bar planes. Neutrons were identified as 
events with no hits in the paddles along the
track to the target, within a
narrow time interval, and in a narrow range of invariant mass $W$ around 
the quasi elastic peak
($\mid W - 938 {\rm MeV} \mid < $50 MeV). 
In addition, cuts on the horizontal
position ($|y_{pos}| < 40$ cm) in the neutron detector and on the angle between $\vec
{q}$ and the neutron momentum ($\theta_{nq} < 110$ mrad) were applied to
optimize the dilution factor. The $\theta_{nq}$ cut   served also to limit the recoil momentum $p_r$ to values
where the model dependence of $A_{ed}^V$ has been shown to be
small (p$_r^{max} \approx$ 85 MeV/c)  \cite{Ar88}.
 The protons were bent vertically in the target field
by nearly 18$^\circ$ almost eliminating their overlap
with the neutrons which further improved their rejection.

\begin{figure}[!ht]
\centerline{\includegraphics[width=3in]{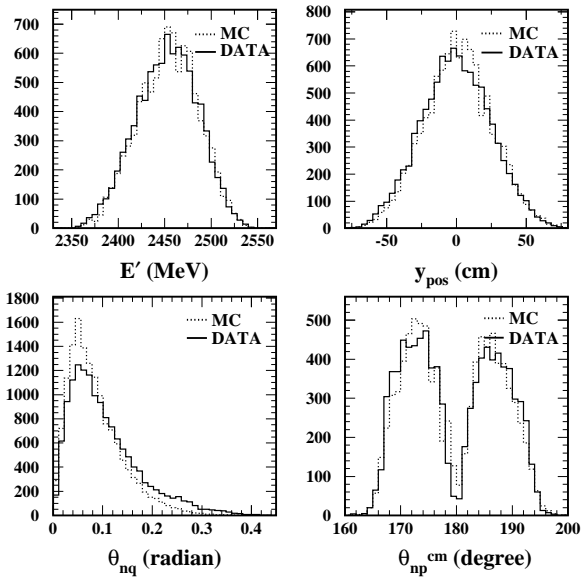}}
\caption{ A comparison between the data and the MC simulation for
(e,e'n) from \nft
for four kinematic variables: $E'$ (scattered electron energy), $y_{\rm pos}$ 
(horizontal position in the neutron detector), $\theta_{nq}$ 
(the angle between $\vec{q}$ and the neutron in the lab), and $\theta_{np}^{cm}$
(the calculated angle between the proton and $\vec{q}$ in the
n-p center of mass).} \label{fig:Monte}
\end{figure}

The experimental asymmetry was diluted by scattering from
materials other than polarized deuterium nuclei. These include
the nitrogen in $^{15}$ND$_3$, the liquid helium, the NMR
coils,  and target  windows. A Monte
Carlo (MC) simulation program was developed \cite{zhu00} to determine the 
dilution factor and to
perform the  detector averaging of the theoretical asymmetries.
It was based on MCEEP \cite{mceep} and included an HMS model, the neutron
detector geometry and approximate efficiencies, the target magnetic
field, the beam raster and radiative effects. Quasi elastic scattering from all
the target materials was simulated in the MC. The normalization was fixed by 
data from carbon (which approximates nitrogen) and liquid helium.
A comparison of the simulated distributions to experimental data is shown in
Fig. \ref{fig:Monte} for four kinematic variables. The good agreement of the distributions
indicates that quasi elastic scattering is the dominant process
for events passing our selection criteria.

The accidental background under the meantime
distribution was 4\% and had no statistically significant asymmetry.
The measured asymmetry was corrected for this dilution.
A correction of 0.2\% was made for the proton contamination. 
Charge exchange in deuterium was taken into
account in the theoretical calculations of final state
interactions (FSI). No correction was applied for charge
exchange reactions with other target materials or the shielding
(estimated to be 0.24\%).
The role of radiative effects on $A_{ed}^V$ was estimated and 
found to be small. Corrections to
the asymmetry for internal radiative effects of 2\%  
and 0.5\% for external effects were applied.

In order to extract $G_E^n$, the corrected experimental asymmetry was
compared to the MC simulation which weights theoretical calculations of the
asymmetry by the event distribution. The theoretical
asymmetries have been calculated on a grid reflecting our
experimental arrangement under different assumptions for the size
of $G_E^n$. Asymmetry values between grid points were obtained by
interpolation.

The theoretical $A^V_{ed}$ values were calculated following
 \cite{Ar88,Ar92}. The calculations are based on a non-relativistic description
of the n--p system in the deuteron, using the Bonn R-Space $NN$
potential  \cite{Ma87} for both the bound state and the description of 
FSI. The full calculations include meson exchange currents and
isobar configurations as well as relativistic corrections. The
dipole parameterization for $G^{n}_{M}$ was
assumed. It was verified that the acceptance averaged value
of $A^V_{ed}$ is linear in the size of  $G^{n}_{M}.$ Thus any
(experimental) value could be incorporated easily if desired. 
 The grid of asymmetries  was calculated for 3 values of
$G_E^n$. In each case the $Q^2$ variation of $G_E^n$ was assumed to be given
by the Galster parameterization \cite{Ga71}(with $p=5.6$) with the magnitude set by an
overall scale parameter of 0.5, 1 or 1.5.
The narrow acceptance in $Q^2$, $0.4 < Q^2 < 0.6$ (GeV/c)$^2$, 
makes the  extracted value of \gen insensitive to the assumed $Q^2$ dependence.

\begin{figure}[!ht] 
\centerline{\includegraphics[width=3.0in]{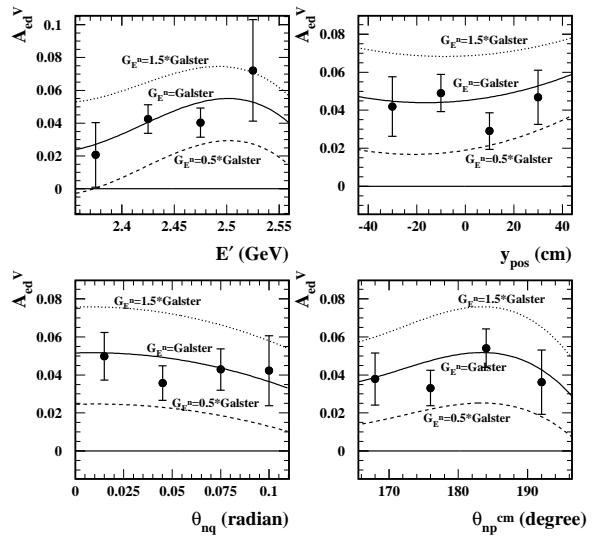}}
 \caption{Comparison between the
measured asymmetries and calculated values of $A^V_{ed}$ for three scaled
parameterizations of Galster shown against four kinematic variables (See Fig.~1).} \label{models} \end{figure}
The MC simulation averaged the asymmetry over all kinematic
variables except the one under investigation. The comparisons are shown in
Fig.~\ref{models}.
To determine the best value of the scale parameter, it was fit as a free
parameter to the data. The resulting value for $G_E^n$ at
$Q^2=0.495$ (GeV/c)$^2$ is $ G_E^n  = 0.04632\pm0.00616 ({\rm stat.}) \pm0.00341 ({\rm syst.})$.

The major sources of systematic errors are:
${\delta P_1^d}/P_1^d = 5.8\%$,
and the uncertainty in determining the average dilution factor is
3.9\%. Cut dependencies give a 2.4\% systematic error. 
Errors associated with the determination of
various kinematic quantities contribute another 2.2\%. 
The determination of $P_e$ contributes 1\%. Finally, the error
in the value of $G_M^n$ was taken to be 1.7\% as derived from a
recent fit to world data  \cite{Jo00}.
The quadratic sum of all the contributions gives a total systematic
error ${\delta G_E^n}/G_E^n = 7.4 \%$.

\begin{figure}[!ht]
\center\includegraphics[angle=90,width=3in]{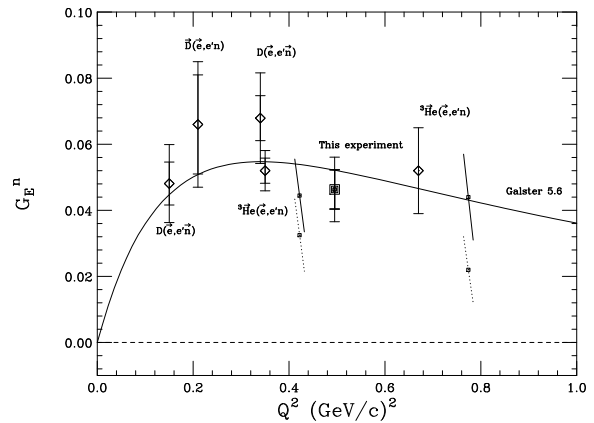}
\caption{Comparison of present experiment with data from recent
polarized scattering
measurements, left to right  \cite{He99}, \cite{Pa99}, \cite{Os99}, \cite{Be99}, \cite{Ro99}. FSI corrections
have been made for all the data except  \cite{Ro99} where
FSI are thought to be modest. Data from Ref.~ \cite{Be99} have been corrected for FSI  \cite{golak00}. The solid line is the
parameterization ($p=5.6$) of Galster  \cite{Ga71}. The 
slanted lines are lattice QCD calculations with (solid lines) and  without (dotted lines)
the disconnected insertions which account for sea-quark effects  \cite{Do98}.} 
\label{pol_dat}
\end{figure}

Our measurement is compared to $G_E^n$ measurements from other
polarized experiments  in Fig.~\ref{pol_dat}.
For reference the standard parameterization of Galster is
shown  \cite{Ga71}. The figure shows also the results of recent
lattice QCD calculations  \cite{Do98}. In these calculations \gen is quite
sensitive to the disconnected insertions which account for the
sea-quarks. The magnitude of these sea contributions to the various
nucleon form factors is almost constant so they are relatively
much more important for $G_E^n$. Thus \gen may provide a valuable
testing ground for lattice calculations of other sea sensitive
quantities such as the strangeness electric and magnetic form factors.

The size of reaction dynamic effects beyond the plane wave Born approximation (PWBA) was determined
by repeating the same extraction procedure using PWBA
calculations. The result for \gen was found to be 13\% smaller
than when it was extracted from $A^V_{ed}$ using the full
calculation. The bulk of the difference is due to FSI.

In conclusion we present the results of a new
measurement of the neutron electric form factor at $Q^2$ = 0.495 (GeV/c)$^2$, 
the highest momentum
transfer to date in polarized scattering using a deuteron target.
This measurement sets a new constraint on the parameterizations of $G_E^n$
and, more importantly, on theoretical models which describe it. In addition it
will
contribute to the extraction of strange quark form factors from parity violating (PV)
elastic scattering from protons \cite{Happex1,Happex2} where errors on previous
measurements of $G_E^n$ are the
largest contributor to the theoretical PV asymmetry, $A_{th}$.  The ongoing effort
to measure  $G_E^n$ will considerably
extend our understanding of  nucleon structure.

We wish to thank the Hall C technical and engineering staffs at TJNAF as well as
the injector, polarized target, FEL and survey groups for their
outstanding support. This work was supported by the Commonwealth of Virginia through
the Institute of Nuclear and Particle Physics at the University of Virginia, 
the Schweizerische Nationalfonds, by DOE contracts
DE-FG02-96ER40950 (University of Virginia), the U.S.-Israel Binational Science Foundation
 (Tel Aviv University) and by Deutsche Forschungsgemeinshaft (SFB 443). 
 The Southeastern Universities Research Association (SURA) operates the
Thomas Jefferson National Accelerator Facility for the United States
Department of Energy under contract DE-AC05-84ER40150.

\end{document}